\documentclass[jcp,showpacs,doublespacing,preprint]{revtex4}
\usepackage{amsfonts}
\usepackage{amsmath}
\usepackage{amssymb}
\usepackage{graphicx}
\usepackage[T1]{fontenc}
\usepackage[latin9]{inputenc}
\usepackage{array}
\usepackage{longtable}
\usepackage{amsmath}
\usepackage{esint}

\begin{document}

\title{The different roles of Pu-oxide overlayers in the hydrogenation of
Pu-metal: An \textit{ab initio} molecular dynamics study based on vdW-DFT+\textit{U}}

\author{Bo Sun}
\thanks{sun\underline{
}bo@iapcm.ac.cn}
\affiliation{Institute of Applied Physics and Computational Mathematics, Beijing 100094,
P.R. China}
\author{Haifeng Liu}
\affiliation{Institute of Applied Physics and Computational Mathematics, Beijing 100094,
P.R. China}
\author{Haifeng Song}
\affiliation{Institute of Applied Physics and Computational Mathematics, Beijing 100094,
P.R. China}
\author{Guangcai Zhang}
\affiliation{Institute of Applied Physics and Computational Mathematics, Beijing 100094,
P.R. China}
\author{Hui Zheng}
\affiliation{Institute of Applied Physics and Computational Mathematics, Beijing 100094,
P.R. China}
\author{Xian-Geng Zhao}
\affiliation{Institute of Applied Physics and Computational Mathematics, Beijing 100094,
P.R. China}
\author{Ping Zhang}
\thanks{zhang\underline{
}ping@iapcm.ac.cn}
\affiliation{Institute of Applied Physics and Computational Mathematics, Beijing 100094,
P.R. China}


\begin{abstract}
 Based on the van der Waals density functional theory
(vdW-DFT)+\textit{U} scheme, we carry out the \textit{ab initio}
molecular dynamics (AIMD) study of the interaction dynamics for
H$_{2}$ impingement against the stoichiometric PuO$_{2}$(111), the
reduced PuO$_{2}$(111), and the stoichiometric $%
\alpha $-Pu$_{2}$O$_{3}$(111) surfaces. The hydrogen molecular
physisorption states, which can not be captured by pure
DFT+\textit{U} method, are obtained by employing the
vdW-DFT+\textit{U} scheme. We show that except for the weak physisorption, PuO%
$_{2}$(111) surfaces are so difficult of access that almost all of
the H$_{2}$ molecules will bounce back to the vacuum when their
initial kinetic energies are not sufficient. Although the
dissociative adsorption of H$_{2}$
on PuO$_{2}$(111) surfaces is found to be very exothermic, the collision-induced dissociation barriers of H$%
_{2}$ are calculated to be as high as $3.2$ eV and $2.0$ eV for
stoichiometric and reduced PuO$_{2}$ surfaces, respectively. Unlike
PuO$_{2}$, our AIMD study directly reveals that the hydrogen
molecules can penetrate into $\alpha $-Pu$_{2}$O$_{3}$(111) surface
and diffuse easily due to the $25$\% native O vacancies located
along the $\langle $111$\rangle $ diagonals of $\alpha
$-Pu$_{2}$O$_{3}$ matrix. By examining the temperature effect and
the internal vibrational excitations of H$_{2}$, we provide a
detailed insight into the interaction dynamics of H$_{2}$ in $\alpha
$-Pu$_{2}$O$_{3}$. The optimum pathways for hydrogen penetration and
diffusion, the corresponding energy barriers ($1.0$ eV and $0.53$
eV, respectively) and rate constants are systematically calculated.
Overall, our study fairly reveals the different interaction
mechanisms between H$_{2}$ and Pu-oxide surfaces, which have strong
implications to the interpretation of experimental observations.
\end{abstract}
\maketitle

\section{Introduction}
Plutonium (Pu) is one kind of strategically vital fissile material
owning attractive nuclear properties for both energy production and
nuclear explosives. From basic point of view, Pu is the most complex
element in the periodic table showing such intricate properties as
the complicated Pu-$5f$ states
\cite{Savrasov2000,Petit2003,Moore2009,Zhu2013}, which are very
sensitive to the slight variations of the surrounding physical and
chemical environments. Thus, since it was first isolated and the
samples became available, Pu has always been challenging the
boundaries of knowledge of the fundamental science and attracting
extensive researching attentions. However, during the
preparation/handling of Pu-sample and the long-term storage of
Pu-device, the reactive Pu metal will easily suffer from the
environment-dependent chemical corrosions, among which the surface
oxidation is to some extent unavoidable because of the ubiquitous
oxygen in the realistic environment. Also, the surface hydrogenation
is catastrophic for Pu and need to be avoided since the Pu-hydride
product can violently catalyze the oxidation (by a factor of
$10$$^{13}$) and even induce the pyrophoricity of Pu. Actually, the
surface corrosion and oxidation of Pu are often treated as
equivalent topics since the readily generated surface layers of Pu
oxides prominently influence other corrosion courses of Pu metal
such as the first step of Pu-hydrogenation named the induction
period. Moreover, given the high radioactivity and toxicity of Pu,
the easily dispersed Pu-hydride and Pu-oxide undoubtedly bring more
hazards to the personnel and the environment during handling and
storage of Pu metal. Therefore, understanding the effects of
Pu-oxide surface layers in the chemical corrosion (especially the
hydrogenation) of Pu metal is crucial for the plutonium science, the
environment, and the nuclear energy sectors. However, it is also
full of challenges that beset experimentalists.

As a matter of fact, the oxidized surface layers are directly
involved in the corrosion process of Pu metal. In many technological
applications of Pu oxides such as in nuclear fuel cycle, it is
always desirable to study the surface properties and surface
reactivity of Pu oxides. Many experiments
\cite{Colmenares1975,Has2000,Butterfield2006} have been devoted to
studying the oxidation mechanisms of Pu metal under various
atmospheres and temperatures, and eventually come to a main
conclusion that the multivalent nature of Pu presents an active
response to the varying environment, in other words, it is difficult
to stabilize Pu at one oxidation state when the surrounding
conditions (such as relative humidity, oxygen content, and the
temperature) change. As a result, the complex Pu-O phase diagram
\cite{Wriedt1990} has not been completely described yet, let along
the detailed surface conditions of such many kinds of Pu oxides.
When exposed to oxygen-rich air, metallic Pu surface oxidizes to a
protective dioxide (PuO$_{2}$) layer
\cite{Has2000,Butterfield2006,Butterfield2004} and a thin
sesquioxide\ (Pu$_{2}$O$_{3}$) layer existing in between. Under
special aqueous
condition, the hydration reaction on PuO$_{2}$ surfaces (PuO$_{2}+x$H$_{2}$O$%
\rightarrow $PuO$_{2+x}+x$H$_{2}$) has been reported to generate
higher compound PuO$_{2+x}$ (x$\leq $$0.27$) \cite{HaschkeSci2000}.
Whereas, it has been theoretically proved to be strongly endothermic
\cite{Korzhavyi2004} and the products (PuO$_{2+x}$) have been
experimentally found to be chemically unstable at elevated
temperatures \cite{Gouder2007}. Thus, PuO$_{2}$ is the stable oxide
as the compound of choice for the long-term storage of Pu and as the
nuclear fuels. However, under oxygen-lean conditions (in the vacuum
or inert gas), the PuO$_{2}$-layer
undergoes the thermodynamically driven auto-reduction (AR) to Pu$_{2}$O$_{3}$%
, including the low-temperature phase of $\alpha $-Pu$_{2}$O$_{3}$
in cubic
structure ($Ia\overline{3}$) and the high-temperature $\beta $-Pu$_{2}$O$%
_{3} $ in a hexagonal structure ($P\overline{3}m1$). More recently,
sub-stoichiometric sesquioxide Pu$_{2}$O$_{3-y}$ ($y=0$ to $1$) has
been obtained from further AR of PuO$_{2}$ layer on Pu surface
\cite{GARC2011}. All of those studies have shown us the complex
surface conditions of oxide-coated Pu, and made us believe that they
are one of the controlling factors for the subsequent corrosion of
Pu. As pointed out hereinbefore, for the safe handling and storage
of Pu-based materials, it is a very important topic to control or
avoid the catastrophic Pu-hydrogenation. Whereas, early experiments
have found that the length of induction period in Pu-hydrogenation
is mainly determined by the component and configuration\ of the
preexisting Pu-oxide overlayer, namely, compared with the PuO$_{2}$
overlayers under ambient conditions, the Pu$_{2}$O$_{3}$ under ultra
high vacuum (UHV) conditions can promote the Pu-hydrogenation
reaction by markedly shortening the induction period \cite{Has2000}.
Since then, experimentalists have invested persistent efforts to
better understand the roles of Pu-oxide overlayers and to quantify
the controlling factors/parameters in detail \cite{MORRALL2007},
aiming to establish a predictive model of Pu-hydrogenation. More
recently, Dinh \textit{et al.} \cite{Dinh2011} have investigated the
role of cubic $\alpha $-Pu$_{2}$O$_{3}$ in the hydrogenation of
PuO$_{2}$-coated Pu by chemically and mechanically altering the
PuO$_{2}$ surface on Pu, and demonstrated that induction periods are
consistently eliminated at conditions that produce catalytic $\alpha $-Pu$%
_{2}$O$_{3}$. They also pointed out that more experimental efforts
are surely needed to clarify such unclear factors for the
unpredictable induction period as the solubility and diffusion of
hydrogen species in Pu oxides. McGillivray \textit{et al.}
\cite{McGillivray2011} have determined hydrogen pressure
($P_{\text{H}_{2}}$) dependence of two early steps (namely,
induction and nucleation) of Pu-hydrogenation reaction by
synthesizing a well reproducible PuO$_{2}$ overlayer. They found
that the induction time and the nucleation rate vary inversely and
linearly with $P_{\text{H}_{2}}$, respectively. From their
standpoint, the diffusion barrier properties of hydrogen in
PuO$_{2}$ are not sensitive to the temperature (cooling from $150$
to $25\,^{\circ}\rm{C}$ ) and the oxygen content (varying from
vacuum to O-rich) during the PuO$_{2}$ synthesis. Although all
experimental studies so far have definitely demonstrated the
distinct roles of PuO$_{2}$ and $\alpha $-Pu$_{2}$O$_{3}$
overlayers, such details as the interaction mechanisms of molecular
H$_{2}$ with Pu-oxide surfaces and the diffusion behaviors of
hydrogen species in Pu oxides have not been clarified, and demand
further study from a very fundamental viewpoint. However, given the
complexity and hazard listed above, it is extraordinarily difficult
to experimentally clarify what happens in the early stages when the
H$_{2}$ molecules are either physisorbed or chemisorbed, and
subsequently the H$_{2}$ either dissociate to create reactive
species or diffuse into the Pu-oxide overlayers. Thus, the atomic
simulations are highly required for the in-depth understanding of
the interaction dynamics of H$_{2}$ molecule with Pu-oxide
overlayers. \

In order to address the above-mentioned issues, theoretical schemes
have to reasonably describe the ground state properties of Pu oxides
at first, and then guarantee that the \textit{ab initio} molecular
dynamics (AIMD) simulations of the interactions between Pu-oxide
surfaces and H$_{2}$ molecules are methodologically reliable and
computationally viable. However, the conventional density-functional
theory (DFT) underestimates the strong on-site Coulomb repulsion of
the $5f$
electrons and incorrectly describes PuO$_{2}$ as a metal \cite%
{Boettger2002} instead of a Mott insulator reported by experiments
\cite{McNeilly1964}.
Fortunately, several beyond-DFT approaches, including the DFT+$U$ \cite%
{Dudarev1998}, the self-interaction corrected LDA
\cite{Petit2003,Petit2010}, the hybrid density functional of (Heyd,
Scuseria, and Enzerhof) HSE \cite{Prodan2005}, and LDA+dynamical
mean-field theory (DMFT) \cite{Yin2008}, have been developed to
correct the failures of conventional DFT in calculations of Pu
oxides. The effective modification of the pure DFT by the DFT+$U$
formalism has been confirmed
widely in studies of physical properties of Pu-oxides (PuO$_{2}$, $\alpha $%
-Pu$_{2}$O$_{3}$ and $\beta $-Pu$_{2}$O$_{3}$) \cite%
{SunJCP2008,SunCPB2008,ZhangP2010,Shi2010,Shi2012,Andersson2009,Jomard2008,Jomard2011,SunJNM2012,SunPLA2012},
such as the insulating ground states, the lattice parameters, the
bulk modulus, the phonon spectra and density of states (DOS), the
surface stability and chemical activity, and the redox energetics.
Note that within the DFT+$U$ framework, both PuO$_{2}$ and
Pu$_{2}$O$_{3}$ are described as the
antiferromagnetic (AFM) insulator. The calculated AFM ground state of Pu$%
_{2} $O$_{3}$ is actually consistent with experimental measurements \cite%
{McCart1981,Wulff1988}. However, unlike the affirmative AFM Pu$_{2}$O$_{3}$ and UO$%
_{2}$, PuO$_{2}$ is peculiar due to its complex ground-state
multiplet, so that strictly speaking its paramagnetic ground state
has not been well understood and an AFM exchange
\cite{Santini1999,Colarieti2002} between Pu ions in PuO$_{2}$ is
still used
to explain the discrepancy between neutron-scattering experiments \cite%
{Kern1999} and magnetic susceptibility measurements
\cite{Raphael1968}, which is beyond the scope of our current work.
In spite of all the published papers and increasing attentions
focusing on Pu oxides, our theoretical understanding on the basic
mechanisms for the hydrogenation of Pu-oxide coated Pu is, to put it
mildly, very poor. Especially, little is known regarding the
interaction behaviors of H$_{2}$ on Pu-oxide surfaces, which is in
sharp contrast to the depth and comprehensiveness of researches
conducted upon the surface reaction mechanisms of ceria
\cite{VICARIO2006,CHEN2007,WATKINS2007,ALAM2011,Yang2010,Paier2013}.
As far as we are aware, Wu \textit{et al.} \cite{Andersson2009,Jomard2008} were the only ones who have studied the interaction between gaseous molecule (H$_{2}$O) and PuO$_{2}$%
(100)/PuO$_{2}$(110) surfaces, whereas in their literature both the
strong correlation effect of Pu-$5f$ electrons and the van der Waals
force between polar H$_{2}$O molecule and PuO$_{2}$ surfaces were
not taken into account. Recently, Jomard and Bottin
\cite{Jomard2011} have via DFT (PBE)+$U$ calculations discussed the
influence of electronic correlations on thermodynamic
stability of PuO$_{2}$ surfaces and pointed out that the O-terminated PuO$%
_{2}$(111) is the most stable surface and DFT+$U$ method can well
describe the electronic structure of the insulating PuO$_{2}$
surfaces, which is a prerequisite to study surface reactivity of
PuO$_{2}$. In order to thoroughly understand the surface properties
of PuO$_{2}$, we have discussed the effects of thickness and
O-vacancy on surface stability and chemical activity of PuO$_{2}$
based on the DFT+$U$ formalism with the spin-orbit coupling (SOC)
effect included, and found that under O-rich conditions, the
stoichiometric O-terminated PuO$_{2}$(111) surface is
thermodynamically stable and chemically inert, while the O-reducing
condition facilitates the formation of on-surface O-vacancy, which
is expected to enhance the chemical activities of PuO$_{2}$(111)
surface \cite{SunJNM2012}.

\begin{figure}[tbp]
\begin{center}
\includegraphics[width=0.8\linewidth]{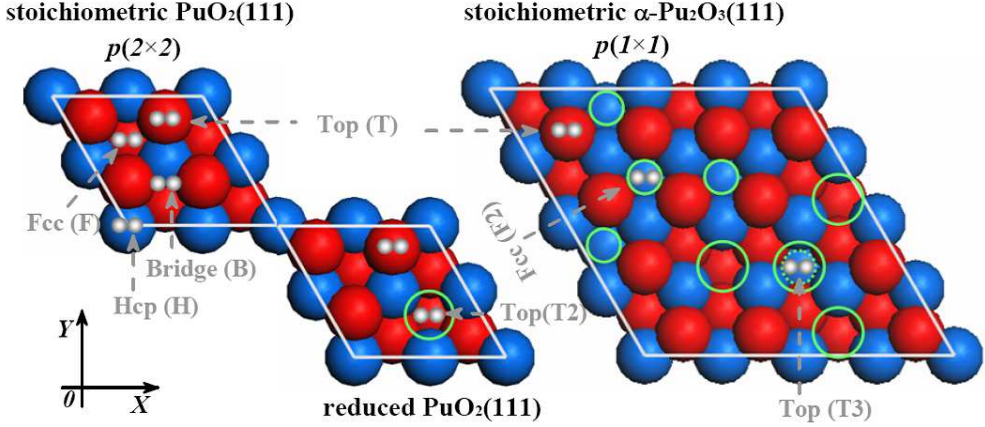}
\end{center}
\caption{(Color online). Top views of Pu-oxide surface
configurations: the $p(2\times 2)$ stoichiometric PuO$_{2}$(111) and
reduced PuO$_{2}$(111) surfaces, and the $%
p(1\times 1)$ stoichiometric $\protect\alpha $-Pu$_{2}$O$_{3}$(111)
surface, where the high-symmetry target sites for H$_{2}$ molecule
are also labeled. The blue, red and dark grey spheres denote Pu, O
and H atoms, respectively. The visible O vacancies on the first, the
second and the third oxygen layers are marked by big solid, small
solid and small dashed green circles, respectively.}
\end{figure}

As a continuation of our theoretical studies on Pu-surface corrosion \cite%
{SunJNM2012,SunPLA2012}, in this paper, we study the interaction
dynamics of H$_{2}$ molecule with Pu-oxide surfaces to clarify the
different roles of PuO$_{2}$ and $\alpha$-Pu$_{2}$O$_{3}$ overlayers
in the hydrogenation of Pu-metal. Specifically, based on the
systematic simulations within the vdW-DFT+$U$ framework, we address
the following issues: (i) the interaction behaviors such as
adsorption, collision, and dissociation of a cold or hot H$_{2}$ on
stoichiometric and reduced PuO$_{2}$(111) surfaces; (ii) the
interaction between H$_{2}$ and stoichiometric $\alpha
$-Pu$_{2}$O$_{3}$(111) surface and the kinetic energy-dependent
penetration of H$_{2}$ into $\alpha $-Pu$_{2}$O$_{3}$(111)
overlayer; (iii) the adsorption and diffusion behaviors of H$_{2}$
in
the $\alpha $-Pu$_{2}$O$_{3}$ matrix, and the interaction between the H$%
_{2}$ of vibrational excitation and the $\alpha $-Pu$_{2}$O$_{3}$
during the diffusion. The rest of this paper is organized as
follows. The details of our calculations are described in Sec. II.
In Sec. III we present and discuss the results. In Sec. IV, we
summarize our main conclusions.

\section{DETAILS OF CALCULATION}

The spin-polarized vdW-DFT+ $U$ calculations are performed using the
Vienna \textit{ab initio} simulation package (VASP)
\cite{Kresse1996} with the projector-augmented wave (PAW) method
\cite{Blo1994} and a plane-wave cutoff
energy of $400$ eV. The Pu-6$s^{2}$7$s^{2}$6$p^{6}$6$d^{2}$5$f^{4}$, O-2$%
s^{2}$2$p^{4}$, and H-1$s^{1}$ electrons are treated as valence
electrons, while the remaining electrons were kept frozen as core
states in the PAW method. The electronic exchange and correlation
are treated within the generalized-gradient approximation (GGA)
using the Perdew-Burke-Ernzerhof (PBE) functional \cite{PERDEW1996},
and the strong on-site Coulomb repulsion among the Pu-5$f$ electrons
is described with the DFT+$U$ scheme formulated by Dudarev $et$
$al.$ \cite{Dudarev1998}. Thus, in this study a Hubbard $U$-like
term ($U_{eff}$=$U\mathtt{-}J$, i.e. the difference between the
Coulomb $U$ and exchange $J$ parameters, hereinafter referred to as
$U$) is added to the PBE functional. In the previous DFT+$U$ studies
of PuO$_{2}$, the choices of $U$ and $J$ parameters are based on an
overall agreement between available experimental data and
theoretical results as regarding the basic physical properties of
the ground state, and the combination of $U$=$4.75$ and $J$=$0.75$
eV has been proved to be reasonable by our previous calculations \cite{SunJCP2008,SunCPB2008,ZhangP2010,Shi2010,SunJNM2012,SunPLA2012} of PuO$_{2}$ and Pu$%
_{2}$O$_{3}$. Furthermore, the DFT+$U$ method has been shown to give
substantially similar results to those obtained by hybrid functionals \cite%
{Prodan2005} with partial exact exchange. Note that the computational demand of DFT+$%
U $ is close to that of conventional LDA or GGA functionals and much
less than that of hybrid functionals, which are computationally too
time-consuming to do the AIMD simulations of such surface systems as
studied in this paper. \

The Pu-oxide surfaces are modeled by finite-sized periodic slab
supercells, consisting of a number of oxide layers infinite in $X$
and $Y$ directions and separated in the $Z$ direction by a vacuum of
$30$ \AA , which is found to sufficiently minimize the interaction
between adjacent slabs. Following our previous DFT+$U$ studies of
PuO$_{2}$ surfaces \cite{SunJNM2012}, in this work we first
calculate the surface energy $E_{\text{s}}^{\text{relax}}$ of the
stoichiometric $\alpha $-Pu$_{2}$O$_{3}$ slabs. This quantity is
written as
\begin{equation}
E_{\text{s}}^{\text{relax}}=\frac{1}{2A}\left( E_{\text{slab}}^{\text{relax}%
}-E_{\text{bulk}}\right) ,
\end{equation}
where $E_{\text{slab}}^{\text{relax}}$ is the total energy of the fully relaxed $%
\alpha $-Pu$_{2}$O$_{3}$ surface slab, $E_{\text{bulk}}$ is the
energy of the reference bulk $\alpha $-Pu$_{2}$O$_{3}$ with the same
number of atoms, and the denominator $2A$ is the total area of both
surfaces of the slab with
a finite thickness. Based on extensive test calculations, the $E_{\text{s}}^{%
\text{relax}}$ of a slab with varying thickness is found to be
converged within $5$ meV/\AA $^{2}$. According to Tasker's general
criteria \cite{Tasker1979} on
surface stability of an ionic crystal, the O-terminated $\alpha $-Pu$_{2}$%
O$_{3}$(111) slab, consisting of successive and electrically neutral
\textquotedblleft O-Pu-O\textquotedblright\ blocks along the $Z$
direction, has the \textquotedblleft
Tasker-type-II\textquotedblright\ surfaces and its surface energy
$E_{\text{s}}^{\text{relax}}$ is calculated to be
$\mathtt{\sim}0.03$ eV/\AA $^{2}$. The $\alpha
$-Pu$_{2}$O$_{3}$(110) is the typical \textquotedblleft
type-I\textquotedblright\ surface stacked with identical neutral
planes
fulfilling the $\alpha $-Pu$_{2}$O$_{3}$ stoichiometry with its $E_{\text{s}%
}^{\text{relax}}$ of $\mathtt{\sim}0.07$ eV/\AA $^{2}$. The polar
\textquotedblleft type-III\textquotedblright\ (001) surface,
consisting of oppositely charged planes, is in this work modeled as
two oxygen-terminated surfaces with 50\% oxygen vacancies to fulfill
the stoichiometric formula and to quench
the dipole moment of the repeated unit in $Z$ direction, and its $E_{\text{s}%
}^{\text{relax}}$ is found to be the largest one
($\mathtt{\sim}0.23$ eV/\AA $^{2}$). Thus, for both PuO$_{2}$ and
$\alpha $-Pu$_{2}$O$_{3}$, the O-terminated (111) surface is most
stable and is briefly named as the (111) surface if not mentioned
differently in this study.

Considering the auto-reduction (AR) of PuO$_{2}$\ to $\alpha $-Pu$_{2}$O$%
_{3} $ driven by the oxygen reducing/poor conditions, in this study,
we therefore mainly study three model surfaces of Pu-oxides,
corresponding to three typical cases of varying oxygen conditions,
namely, stoichiometric/ideal PuO$_{2}$(111) surface (oxygen rich
condition), reduced/defective PuO$_{2}$(111) surface with $25$\% O
vacancies (oxygen reducing), and stoichiometric $\alpha
$-Pu$_{2}$O$_{3}$(111) surface (oxygen poor). The top views of
$p(2\times 2)$ stoichiometric and reduced PuO$_{2}$(111) surfaces
and $p(1\times 1)$ stoichiometric $\alpha $-Pu$_{2}$O$_{3}$(111)
surface used in the calculations are shown in Fig. 1. The
Brillouin zone (BZ) integration is performed using the Monkhorst-Pack (MP) $%
k $-point mesh \cite{Monkhorst1976}. Specifically, for both
stoichiometric and reduced PuO$_{2}$(111) surfaces, the
two-dimensional (2D) $p(2\times 2)$ cell with a $7\times 7\times 1$
$k$-point MP grid is employed in the static calculations, and a
$3\times 3\times 1$ $k$-point MP grid is used in AIMD simulations.
Since the ideal
stoichiometric $\alpha $-Pu$_{2} $O$_{3}$ is structurally similar to the 2$%
\times $2$\times $2 fluorite PuO$_{2}$\ ($Fm\overline{3}m$)
supercell containing $25$\% O vacancy located in the 16$c$
($0.25,0.25,0.25$) sites, the surface area of unit $p(1\times 1)$
$\alpha $-Pu$_{2}$O$_{3}$(111) cell is equal to that of the
$p(4\times 4)$ PuO$_{2}$(111) cell. Thus, a $3\times 3\times 1$
$k$-point MP grid is used in static calculations and it is
restricted to the $\Gamma $-point in AIMD simulations. The slab of $%
p(2\times 2)$-PuO$_{2}$(111) consists of $6$ \textquotedblleft
O--Pu--O\textquotedblright\ blocks along the $Z$ direction,
including
totally $24$ plutonium and $48$ oxygen atoms, and the $\alpha $-Pu$_{2}$O$%
_{3}$(111) surface model contains $4$ \textquotedblleft
O--Pu--O\textquotedblright\ blocks with $64$ Pu and $96$ O atoms in
the supercell. In the spin-polarized static calculations of Pu-oxide
slabs with
the AFM orders of the Pu-sublattice set to be in a simple \textquotedblleft $%
\uparrow $ $\downarrow $ $\uparrow $ $\downarrow
$\textquotedblright\ alternative manner along the $Z$ direction, the
two \textquotedblleft O--Pu--O\textquotedblright\ oxide bottom\
blocks are fixed to their
calculated bulk positions ($a_{\text{PuO}_{\text{2}}}=5.466$ \AA\ and $%
a_{\alpha \text{-Pu}_{\text{2}}\text{O}_{\text{3}}}=11.20$ \AA ) and
all other atoms are fully relaxed until the Hellmann-Feynman forces
become less than $0.01$ eV/\AA . The dipole-dipole interactions
along the direction perpendicular to the slab are also corrected, as
implemented in the VASP code \cite{Makov1995,Neugebauer1992}.

Due to the \textquotedblleft Tasker-type-II\textquotedblright\
character of PuO$_{2}$(111) and $\alpha $-Pu$_{2}$O$_{3}$(111)
surfaces, the non-coplanar O-anion and Pu-cation may generate
surface dipole moment along the $Z$ direction, especially when the
surface O-vacancies exist. In order to detailedly discuss the
multiple interaction mechanisms between H$_{2}$ and Pu-oxide
surfaces, van der Waals (vdW) dispersion forces upon H$_{2}$ had
better be considered. In this study, we employ the non-local
vdW-DFT+$U$ approach \cite{Dion2004,Klime2010} to
calculate the static potential-energy pathway (PEP) along which a cold H$%
_{2} $ approaches to Pu-oxide surfaces, and to simulate the dynamic
interaction between a hot H$_{2}$ (with certain initial kinetic
energy) and the Pu-oxide surfaces. Moreover, we expect that the
static PEP calculations and the AIMD simulations can complement each
other to achieve a thorough understanding of the energetics factor
and dynamics process determining the distinct roles of PuO$_{2}$ and
$\alpha $-Pu$_{2}$O$_{3}$ overlayers in Pu hydrogenation. From the
PEP results, we can analyze the binding strength between Pu-oxide
and H$_{2}$ molecule via the adsorption energy $E_{\text{ad}}$,
which is given by
\begin{equation}
E_{\text{ad}}=E_{\text{H}_{\text{2}}\text{/Pu-oxide}}-\left( E_{\text{%
Pu-oxide}}+E_{\text{H}_{\text{2}}}\right) .
\end{equation}
Here, $E_{\text{H}_{\text{2}}\text{/Pu-oxide}}$, $E_{\text{Pu-oxide}}$ and $%
E_{\text{H}_{\text{2}}}$ are the total energies of the adsorption
system, the pure Pu-oxide system, and the H$_{2}$ molecule without
zero-point vibration, respectively.

The AIMD simulations under the Born-Oppenheimer approximation are
performed using the Verlet algorithm with a time step of $0.5$ fs
within the micro
canonical ensemble (NVE). With the NVE-AIMD, the very fast impingement of H$%
_{2}$ molecule against the Pu-oxide surfaces is simulated to uncover
the possible molecular dissociation and estimate the high
dissociation
barriers. Based on the harmonic approximation and the vibrational energy $E_{%
\text{vib}}$ written as
\begin{equation}
E_{\text{vib}}=(\tfrac{1}{2}+n)h\nu,
\end{equation}
the ground-state zero-point vibration ($n=0$) and internal
vibrational excitation ($n=1$ or $2$) of H$_{2}$ molecule with a
calculated harmonic frequency ($\nu=128$ THz) are also considered.
In addition, the canonical ensemble (NVT) at several prefixed
temperatures through a Nos\'{e} thermostat \cite{Nose1984} has also
been employed as an alternative approach to discuss the high
temperature effect on the state or behavior of H$_{2}$ molecule in
$\alpha $-Pu$_{2}$O$_{3}$. However, the short time scale that can be
simulated by the AIMD based on vdW-DFT+$U$ is a major limitation for
such rare events as the penetration and diffusion of H$_{2}$
molecule in this study, which usually take place on a much longer
timescale (of milliseconds or more) than the atomic vibrational
timescale of femtoseconds. Here, we attempt to depict such rare
events by\ using the harmonic transition state theory (hTST) with
the rate constant given by
\begin{equation}
k_{\text{hTST}}^{\text{Pen/Diff}}=\frac{\Pi
_{i}^{3N}\nu_{i}^{\text{IS}}}{\Pi
_{i}^{3N-1}\nu_{i}^{\text{TS}}}e^{-\Delta E/k_{B}T},
\end{equation}
where $\nu_{i}^{\text{IS}}$ and $\nu_{i}^{\text{TS}}$ are the normal
mode vibrational frequencies of the initial and transition states,
respectively. Here, the energy barrier $\Delta E$ is defined as the
energy difference between the initial state (IS) and transition
state (TS), where the TS structures for relevant processes are
located by employing the climbing image nudged elastic band (CI-NEB)
method \cite{HENKELMAN2000} using at least three images along each
pathway. The vibrational frequencies are obtained by diagonalization
of the force-constant matrix in Cartesian coordinates (Hessian),
where the force constants are obtained from finite differences of
the forces with atomic displacements of $\pm 0.02$ \AA . Then, we
can further use the frequencies to characterize whether an optimized
stationary point is an initial/minimum state without imaginary
frequency or a TS with only one imaginary frequency.

\section{Results and discussion}

Since the surface interaction with molecular monomer is very
site-specific, as shown in Fig. 1, in this study we consider four
high-symmetry target sites (T1, H1, F1, and B1) for a H$_{2}$
molecule approaching the stoichiometric PuO$_{2}$(111) surface, and
also four sites (T1, T2, H1, and H2) on the reduced PuO$_{2}$(111)
surface with an artificial surface oxygen vacancy marked by a solid
green circle in Fig. 1. For the stoichiometric $\alpha
$-Pu$_{2}$O$_{3}$(111) surface shown in Fig. 1, where the visible O
vacancies on the first/outmost, the second/subsurface, and the
third/deep oxygen layers are marked by big solid, small solid, and
small dashed green circles, respectively, we consider three
adsorption sites (T1, T3, and F2). Due to the $25$\% O vacancies
existing inherently in $\alpha $-Pu$_{2}$O$_{3}$,\ one can see that
T3 is actually a special site, under which two superposed O
vacancies seem to bore a hole in the $\alpha $-Pu$_{2}$O$_{3}$(111)
slab and may act as an entrance for the smallest diatomic molecule
H$_{2}$ with its bond-length of only $0.75$ \AA .

The H$_{2}$ molecule is initially placed over different Pu-oxide
surface sites with a height of $4$ or $4.5$ \AA\ and the initial
orientation at each site is set to be along the $X$, $Y$, and $Z$
axes, respectively. Altogether, we construct $33$ high-symmetry
initial geometries of H$_{2}$ on three Pu-oxide surfaces and name
them by adding a corresponding postfix---T1-$X$ for example, among
which $11$ `-$X$' structures are presented in Fig. 1. In
addition, we also generate some low-symmetry initial geometries by rotating H%
$_{2}$ with small angles, and find that low-symmetry geometries can
relax into the high-symmetry ones after geometry optimization. Thus,
in the static PEP calculations, the H-H bond is fixed to be either
parellel or perpendicular (i.e., the `-$Z$' cases) to the (111)
surface when the H$_{2}$ molecule drops step by step from its
initial location. Since a H$_{2}$ molecule on a solid surface has
$6$ degrees of freedom ($6$-DOF) [see the inset to Fig. 3(e), i.e.,
bond-length $d$, height $h$, azimuth angles ($\theta $ and $\phi $) of H$%
_{2} $, and projected positions ($X$, $Y$) of mass center], so that
in the
static PEP calculations we consider in practice $5$-DOF of H$_{2}$ except for $%
\theta $. However,\ during the AIMD simulations of the H$_{2}$
molecule with a given kinetic energy $E_{\text{k}}^{\bot }$
impingement against the initially resting/cold Pu-oxide surface, we
let both the molecule and the surface layers relax freely.

\subsection{The interaction between H$_{2}$ and PuO$_{2}$(111) surfaces}
In this subsection, we focus on the interaction mechanisms between
H$_{2}$ and PuO$_{2}$(111) surfaces, namely, stoichiometric/ideal
PuO$_{2}$(111) and reduced/defective PuO$_{2}$(111) surfaces (see
Fig. 1).

\begin{figure}[tbp]
\begin{center}
\includegraphics[width=0.8\linewidth]{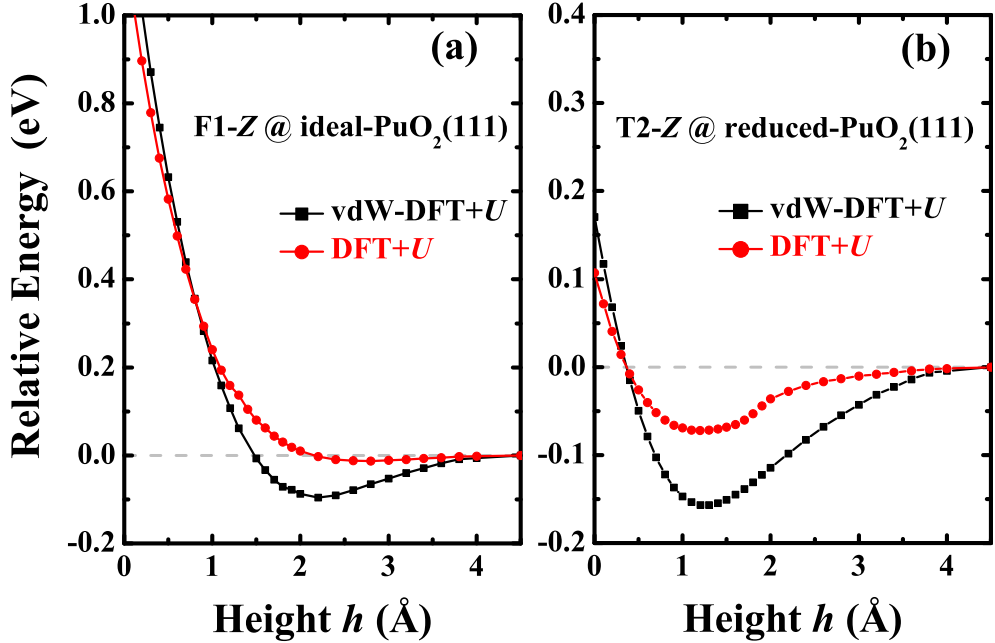}
\end{center}
\caption{(Color online) Calculated PEPs of the
H$_{2}$ molecule approaching (a) the ideal-PuO$_{2}$(111) surface from the initial F1-%
$Z$ geometry; and (b) the reduced-PuO$_{2}$(111) from T2-$Z$. The
initial geometries with $h_{0}$ = $4.5$ \r{A} are considered as the
reference systems, with their total energies set to 0 eV. }
\end{figure}

First of all, the effects of vdW forces upon H$_{2}$ molecule from
the ideal-PuO$_{2}$(111) surface are probed by an overall comparison
between vdW-DFT+$U$ and DFT+$U$ calculated PEPs. We find that the pure DFT+$%
U$ calculations usually can not capture the physisorption state of
H$_{2}$ (with the adsorption energy $E_{\text{ad}}$ of H$_{2}$
underestimated to less than $15$ meV); not but what the vdW-DFT+$U$
scheme is able to do that. The improved description
of H$_{2}$ physisorption by vdW-DFT+$U$ is particularly so when diatomic molecule H$%
_{2}$ is perpendicular to the ideal-PuO$_{2}$(111) surface, for
example, the F1-$Z$ PEPs presented in Fig. 2(a). When H$_{2}$ drops
from F1-$Z$ of $h_{0}$ = $4.5$ \AA, the DFT+$U$ PEP indicates nearly
an onefold force of repulsion existing between H$_{2}$ molecule and
ideal-PuO$_{2}$(111) surface, since the variation of relative energy
as a function of $h$ can be well fitted by $1/h$%
. However, the vdW-DFT+$U$ PEP shows the twofold interaction of both
repulsion and attraction between H$_{2}$ and ideal-PuO$_{2}$(111)
surface,
and thus the physisorption state of H$_{2}$ ($h$ $\approx $ $2.2$ \AA\ and $%
E_{\text{ad}}\approx $ $-0.10$ eV) is well discerned mainly due to
the vdW attractive force upon the H$_{2}$ from the `Tasker-type-II'
PuO$_{2}$(111) surface, which is not entirely considered in the pure
DFT+$U$ method. We note that the DFT+$U$ calculated $E_{\text{ad}}$
of H$_{2}$ with the T1-$Z$ geometry on the ideal-CeO$_{2}$(111)
surface is only $\mathtt{\sim}20$ meV \cite{CHEN2007}, which is also
notably underestimated. Since the surface O-vacancy can enhance the
chemical activity of the inert ideal-PuO$_{2}$(111) surface by
efficiently lowering the surface work function \cite{SunJNM2012},
some DFT+$U$ PEPs seem to be able to give the weak
physisorption states of H$_{2}$ on reduced-PuO$_{2}$(111) with $E_{\text{ad}%
}\ $of only $-50$ meV more or less, but even so they can not really
describe the interaction between H$_{2}$ and reduced-PuO$_{2}$(111)
because of the missing vdW force in the treatment. Figure 2(b) shows
as an example the T2-$Z$ PEP at reduced-PuO$_{2}$(111). As a matter
of fact, the surface O-vacancy will induce the local polarization of
ideal-PuO$_{2}$(111) surface and produce a stronger vdW attraction
for H$_{2}$. We can see from Fig. 2(b) that the vdW-DFT+$U$
calculated physisorption energy $E_{\text{ad}}$ of H$_{2}$ near the
O-vacancy (namely, $h$ $\approx $ $1.3$ \AA\ over T2-$Z$ site) on
reduced PuO$_{2}$(111) surface is about $-0.17$ eV instead of $-50$
meV from pure DFT+$U$, and is somewhat higher of $\mathtt{\sim}70$
meV than the cases of H$_{2}$ on ideal-PuO$_{2}$(111) such as F1-$Z$
in Fig. 2(a). Therefore, in order to reasonably describe the
multiple interaction between H$_{2}$ and Pu-oxide surfaces, we
choose the vdW-DFT+$U$ scheme instead of pure DFT+$U$ method in the
following static PEP calculations and AIMD simulations.

\begin{figure*}[tbp]
\begin{center}
\includegraphics[width=0.8\linewidth]{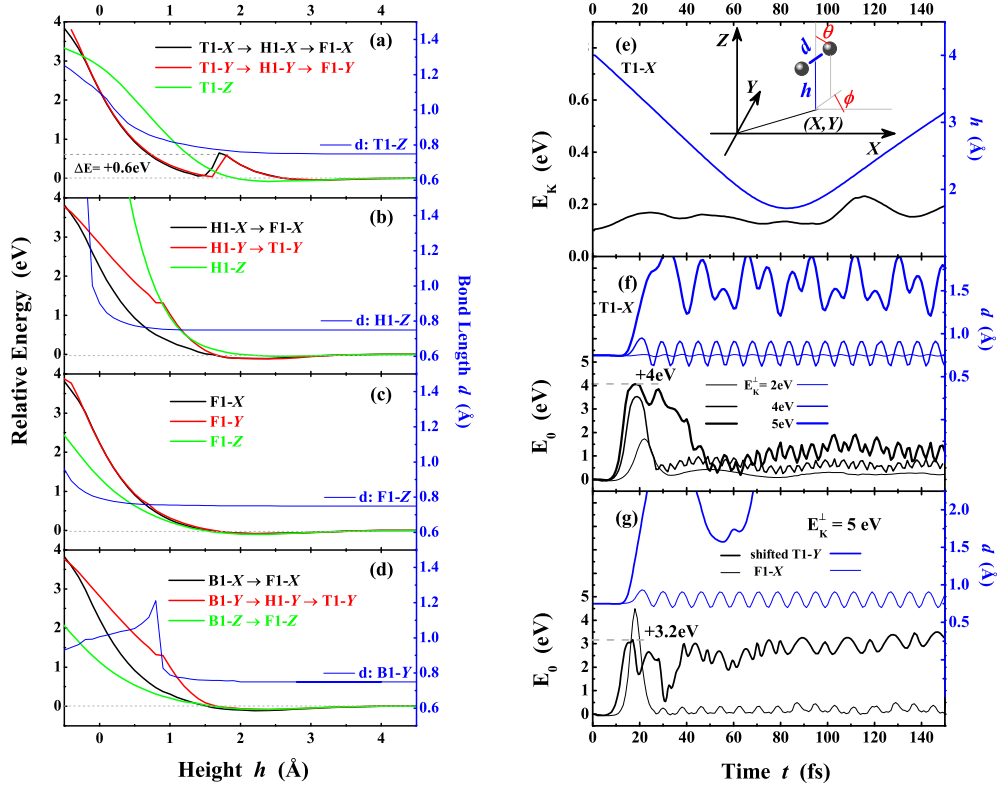}
\end{center}
\caption{(Color online) Left column: Static PEPs and bond-length $d$
(in \r{A}) of a H$_{2}$ molecule approaching the four different
sites [namely, (a) T1, (b) H1, (c) F1, and (d) B1] on
ideal-PuO$_{2}$(111) surface from different initial geometries, and
the approximate directions of H$_{2}$ lateral migrations are
indicated by arrows. Right column: AIMD results of
the interaction dynamics for a H$_{2}$ impingement against ideal-PuO$_{2}$%
(111) surface. The inset shows the six degrees of freedom for
H$_{2}$ on surface. When started from T1-$X$, the evolutions of (e)
kinetic energy $E_{\text{k}}$ (in eV) of the system and height $h$
(in \r{A}) of the H$_{2}$ molecule with
initial E$_{\text{k}}^{\perp }$ of $0.1$ eV and (f) electronic free-energy $E%
_{\text{0}}$ of the system and $d$ of H$_{2}$ for the increasing $E_{\text{k}%
}^{\perp }$. (g) Two limiting cases for the collision induced
dissociation of H$_{2}$ with $E_{\text{k}}^{\perp }$ of $5$ eV,
namely, started from shifted T1-$Y$ and F1-$X$ geometries. }
\end{figure*}

Figure 3(a)-(d) present all the vdW-DFT+$U$ PEPs of H$_{2}$
approaching the ideal-PuO$_{2}$(111) surface considered in this
work. One can see
that H$_{2}$ molecules first undergo a weak physisorption state with $E_{%
\text{ad}}$ $\geq $ $-0.10$ eV when $h$ $\geq $ $2.0$ \AA, and then
it seems to be hard for H$_{2}$ to get further close ($h$ $\leq $
$2.0$
\AA ) to the inert surface, as the dominant force of repulsion between H$%
_{2}$ and ideal-PuO$_{2}$(111) increases rapidly and notably raises
the total energy of these systems, especially when $h$ $\leq $ $1.0$
\AA. We also find that many H$_{2}$ molecules can diverge away from
their initial pathways driven by the lateral components of the
repulsive force. These surface migrations of H$_{2}$ are labeled by
their approximate directions in Figs. 3(a), 3(b), and 3(d). Much to
our surprise, in Fig. 3(a) H$_{2}$ molecules with the T1-$X$ and
T1-$Y$\ initial geometries can first overcome an energy barrier of
$\mathtt{\sim}0.6$ eV at $h\approx $ $1.7$ \AA\ to migrate to H1
site, and then move towards F1 site. Among all these PEPs with $h$
ranging from $4.5$ to $-0.5$ \AA , none but the H1-$Z$ PEP presents
the
dissociative adsorption of H$_{2}$ molecule when $h$ $\leq $ $-0.1$ \AA\ %
[see Fig. 3(b)], with one H atom binding to the subsurface Pu-cation
and the other leaving over the surface, and the corresponding energy
barrier is much
higher than $4.0$ eV, which indicates that the dissociative adsorption of H$%
_{2}$ on ideal-PuO$_{2}$(111) surface should be an extremely
endothermic and rare event. However, after artificially breaking the
H-H bond, the adsorption energy $E_{\text{ad}}$ of H atoms binding
to the surface O anions is calculated to be $-2.3$ eV, which is
really a very exothermic reaction to hydroxylate the PuO$_{2}$(111)
surface with H atoms. So that the dissociative adsorption should be
the most stable and favorable state of H$_{2}$ molecule on
ideal-PuO$_{2}$(111) surface, which is also found to be similar to
the strongly exothermic dissociation of H$_{2}$ on the
ideal-CeO$_{2}$(111) surface \cite{VICARIO2006,CHEN2007,WATKINS2007}
with the $E_{\text{ad}}$ of H atoms calculated by the DFT+$U$ to be
about $-2.6$ eV. And the two OH groups formed on the oxide surface
can help to reduce the nearby Pu$^{4+}$/Ce$^{4+}$ cations to $+3$
state.

Given the existing high barrier, the most controversial issue about
the dissociative adsorption of H$_{2}$ actually rests with the
reliable method to effectively simulate this extremely rare event
and reasonably estimate the corresponding dissociation barriers.
Recently, Alam \textit{et al.} \cite{ALAM2011} have simulated the
interaction process of hot H$_{2}$ molecule with the
ideal-CeO$_{2}$(111) and (110) surfaces by using the ultra
accelerated quantum chemical molecular dynamic (UA-QCMD) method,
which is based on the tight-binding quantum chemistry and the
classical MD programs. In their UA-QCMD simulations, the mean
velocity of several km/s was given to the H$_{2}$ molecule. Here, we
perform comprehensive NVE-AIMD simulations to fulfil this task. From
Fig. 3(e) we can see that if its initial kinetic energy is set at
$0.1$ eV, the initially nonrotating and nonvibrating H$_{2}$ from
T1-$X$ of $h_{0}$ = $4.0$ \AA  \ will bounce
back to the vacuum soon and even can not touch the ideal-PuO$_{2}$%
(111) surface. When the initial $E_{\text{k}}^{\bot }$ of H$_{2}$ is
less than $1.0$ eV, the diffraction behaviors of molecular H$_{2}$
seem to be similar to the case presented in Fig. 3(e) without
reference to its initial geometries. By continually elevating the
initial $E_{\text{k}}^{\bot }$ of H$_{2}$ up to $5.0$ eV,
we find that most of H$_{2}$ molecules parallelly hitting the ideal-PuO$_{2}$%
(111) surface can be broken into two hot H atoms owning high
velocities, which are expected to be captured by the surface O
anions and eventually form two surface hydroxyl groups, i.e., the
above-mentioned stable dissociative adsorption state of H$_{2}$.
Whereas, among all AIMD
trajectories, the $E_{\text{k}}^{\bot }$-dependent dissociation behavior of H$%
_{2}$ from T1-$X$ is an interesting exception and is thus presented
in Fig. 3(f). It shows in Fig. 3(f) that (i) when the initial
$E_{\text{k}}^{\bot }=2.0$ eV, H$_{2}$ can not touch the surface
O-anion and bounce straightly without arousing the molecular
zero-point vibration, according to the inset of Fig. 7(b); (ii) when
its $E_{\text{k}}^{\bot }$ reaches $4.0$ eV, the internal
vibrational excitation of H$_{2}$ molecule emerges [$n=1$ in Eq.
(3)] after the head-on collision with the O-anion; (iii) until
$E_{\text{k}}^{\bot }$ reaches $5.0$ eV, molecular H$_{2}$ is
dissociated by its violent collision with surface O-anion. However,
the products are not two hot H atoms but a hot H$_{2}$O molecule
released to the vacuum and an O-vacancy left on the surface. This
surface reduction process has also been reported by the
above-mentioned UA-QCMD simulations of high-energy colliding H$_{2}$
molecule on CeO$_{2}$(111) and (110) surfaces \cite{ALAM2011}. The
energy barrier of collision induced dissociation process in this
study is estimated to be $\mathtt{\sim}4.0$ eV based on the
electronic free energy $E_{0}$ of this system in Fig. 3(f), which is
close to the bond energy ($\mathtt{\sim}4.5$ eV) of a free-standing
H$_{2}$ molecule and indicates that
the inert ideal-PuO$_{2}$(111) surface can not act as the catalyst for H$%
_{2} $ dissociation. Since moisture can strongly enhance the further
oxidation of Pu metal \cite{Colmenares1975,Has2000}, one can see
that both the surface hydroxylation of PuO$_{2}$ by H$_{2}$ and the
interaction between H$_{2}$O and PuO$_{2}$ surface are important
topics, which will be investigated in our next work. In order to
search out the possible lowest dissociation barrier, different
incident conditions (namely, the initial $E_{\text{k}}^{\bot }$ and
the angles of incidence in some trajectories) are considered. Figure
3(g) presents two limiting cases. When $E_{\text{k}}^{\bot }$ is
$5.0$ eV, the AIMD trajectory of H$_{2}$ started from the
horizontally shifted T1-$Y$ geometry (with one H-atom right above
the surface O-anion) turns out to be the most possible dissociative
adsorption path with the lowest barrier of $\mathtt{\sim}3.2$ eV.
However, the trajectory started from F1-$X$ indicates that H$_{2}$
will never dissociate even if we would further elevate the
$E_{\text{k}}^{\bot }$ in the NVE-AIMD simulation. Thus, the
collision-dissociation behavior of H$_{2}$ is very site specific on ideal-PuO%
$_{2}$(111) surface and depends heavily on the incident condition of H$_{2}$%
. Overall, all the AIMD results well support the static PEPs and
leads to a conclusion that the ideal-PuO$_{2}$(111) surface is
really difficult of access for H$_{2}$ molecule.

\begin{figure*}[tbp]
\begin{center}
\includegraphics[width=0.8\linewidth]{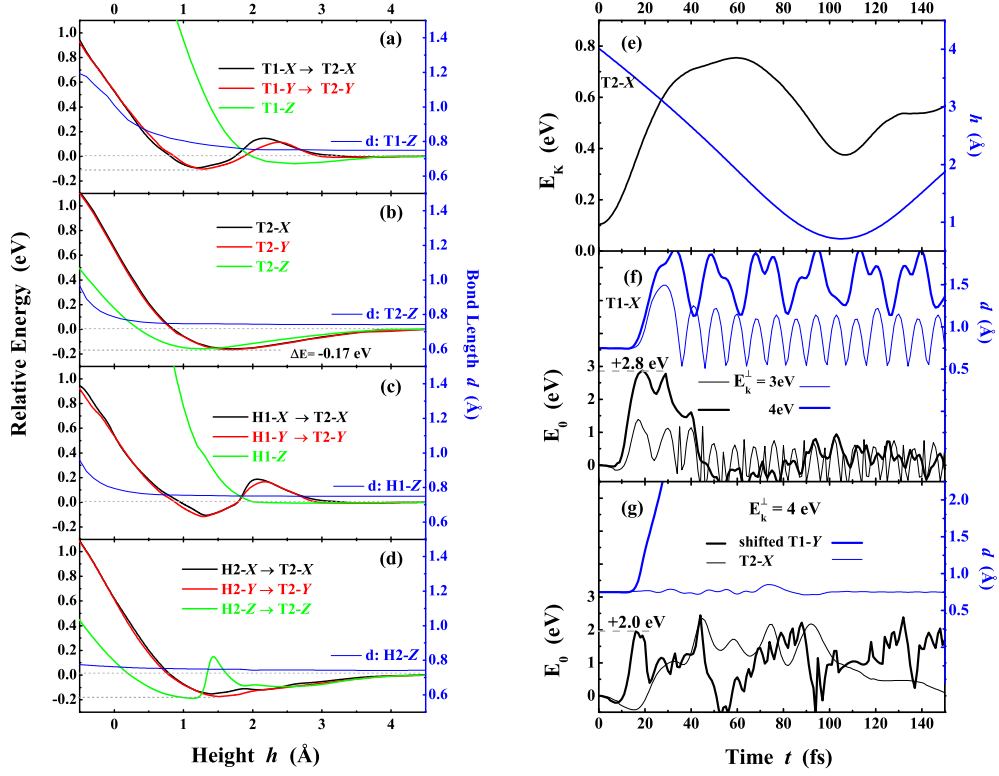}
\end{center}
\caption{(Color online) Left column: Static PEPs and $d$ of a
H$_{2}$ molecule approaching the four target sites [namely, (a) T1,
(b) T2, (c) H1, and (d) H2] on reduced-PuO$_{2}$(111) surface from
different initial geometries. Right column: AIMD results of
H$_{2}$ interaction with reduced-PuO$_{2}$(111) surface. (e) Evolutions of $E_{\text{k}}$ of the system and $h$ of the H$%
_{2}$ molecule with initial geometry of T2-$X$ and
$E_{\text{k}}^{\perp }$ of $0.1$ eV. (f) Evolutions of
$E_{\text{0}}$ of the system and $d$ of the H$_{2}$ molecule started
from T1-$X$ with increasing $E_{\text{k}}^{\perp }$. (g) Two
limiting cases for the collision induced dissociation of H$_{2}$
with $E_{\text{k}}^{\perp }$ of $4$ eV, i.e., started from shifted
T1-$Y$ and T2-$X$ geometries. }
\end{figure*}

As the format of Fig. 3, Fig. 4 presents the interaction behaviors
of H$_{2}$ molecule with the reduced-PuO$_{2}$(111) surface, based
on all static PEP results and some representative AIMD trajectories
plotted in the left- and right-hand columns, respectively. From the
vdW-DFT+$U$ PEPs, we can see that
the molecular physisorption states are ubiquitous for H$_{2}$ on reduced-PuO$%
_{2}$(111) surface when $h$ $\geq $ $1.3$ \AA , and the
physisorption states of H$_{2}$ over the surface O-vacancy (i.e., T2
site) seem to be more
stable than others, so that when further depressing the nearby adsorbed H$%
_{2}$ molecules, most of them tend to stride over certain barriers
(less than $0.3$ eV) to migrate to T2 site. Note that H$_{2}$ seems
to get much closer to the reduced-PuO$_{2}$(111) surface than to the
ideal-PuO$_{2}$(111) surface mainly because of the 25\% O-vacancy on
the former.
However, when $h$ $\leq $ $1.3$ \AA, it becomes more and more difficult for H%
$_{2}$ to get further closer to the surface and all these PEPs with
$h$ ranging from $4.5$ to $-0.5$ \AA\ do not present the
dissociative adsorption of H$_{2}$, which is to some extent similar
to the behavior of H$_{2}$ on ideal-PuO$_{2}$(111) surface.

Figure 4(e) shows the total kinetic energy $E_{\text{k}}$ of the system and the bond-length $%
d$ of H$_{2}$ with its initial $E_{\text{k}}^{\bot }$ of $0.1$ eV
along\ AIMD trajectory started from T2-$X$ of $h_{0}$ = $4.0$ \AA .
We can see that during the initial $60$ fs, the vdW force of
attraction causes the steady augmentation of $E_{\text{k}}$, and
then the increasing force of repulsion exceeds the vdW force and
pushes H$_{2}$ back to the vacuum. Meanwhile, the zero-point
vibration of H$_{2}$ is aroused. Comparing Fig. 3(e) with Fig. 4(e),
we can see that the interaction of H$_{2}$ with ideal- and
reduced-PuO$_{2}$(111) surfaces are actually unalike, and the
surface
O-vacancy modulates the interaction behavior between H$%
_{2}$ and PuO$_{2}$(111) surfaces by changing the surface atomic and
electronic structure properties. Based on enough NVE-AIMD
trajectories of H$_{2}$ impinging against the reduced-PuO$_{2}$(111)
surface, we also try to reveal the possible dissociation behavior
and search out the lowest dissociation barrier of H$_{2}$ on
reduced-PuO$_{2}$(111) surface. We find that several contact
collisions between H$_{2}$ and reduced-PuO$_{2}$(111) can not occur
until the initial $E_{\text{k}}^{\bot }$ of H$_{2}$ reaches $1.0$
eV, which is consistent with most of the PEPs in the left-hand
column of Fig. 4. When the $E_{\text{k}}^{\bot }$ of H$_{2}$ reaches
$4.0$ eV, the collision induced dissociation of H$_{2}$ will take
place to yield two hot H atoms in most AIMD trajectories. Whereas,
the trajectory started from T1-$X$ still make an interesting
exception, which is revealed in Fig. 4(f). From Fig. 4(f), we can
see that when $E_{\text{k}}^{\bot }=3.0$ eV, after the quick
migration [from T1 to T2 site driven by the repulsive force, see
Fig. 4(a)] and the subsequent collision (at T2 site), the rebounded
H$_{2}$ molecule seems to be at the second vibrational excitation
state [$n=2$ in Eq. (3)]. When $E_{\text{k}}^{\bot }$ reaches $4.0$
eV, without the lateral migration, the H$_{2}$ molecule straight
hits the O-anion target to break its bond, and
soon two hot H atoms pull the O-anion out of the surface to yield a hot H$%
_{2}$O molecule released to the vacuum. Upon that, when starting
from T1-$X$, the collision-dissociation dynamics of H$_{2}$ on
reduced-PuO$_{2}$(111) surface is similar to that occurred on the
ideal-PuO$_{2}$(111) surface. However, the corresponding
dissociation barriers are quite different, namely, $2.8$ and $4.0$
eV for H$_{2}$ on reduced- and ideal-PuO$_{2}$(111) surfaces,
respectively. For the dissociation possibility of H$_{2}$ (with
$E_{\text{k}}^{\bot }=4.0$ eV) on reduced-PuO$_{2}$(111) surface,
Fig. 4(g) presents two limiting cases as follows: (i) The AIMD
trajectory started from the horizontally shifted T1-$Y$ geometry [as
in Fig. 3(g)] gives the lowest energy barrier of $\mathtt{\sim}2.0$
eV, much lower than the lowest barrier $\mathtt{\sim}3.2$ eV in Fig.
3(g); (ii) Whereas, the trajectory started from T2-$X$ (or T2-$Y$,
not shown here) indicates that H$_{2}$ will never dissociate even
with a higher $E_{\text{k}}^{\bot }$. Given the incidental
migrations of H$_{2}$ from other sites to T2 (see the PEP results in
Fig. 4), we can conclude that the reduced-PuO$_{2}$(111) surface is
also difficult of access for H$_{2}$ molecule, let along the
collision induced dissociation.

\begin{figure*}[tbp]
\begin{center}
\includegraphics[width=0.7\linewidth]{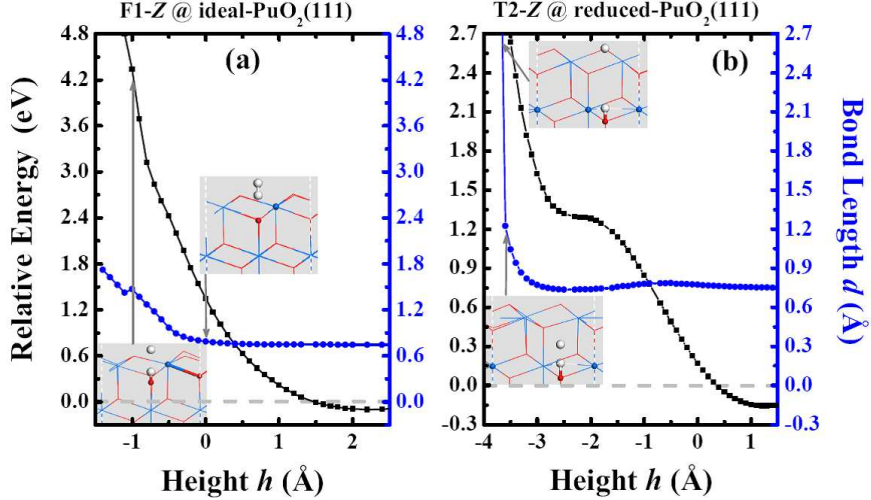}
\end{center}
\caption{(Color online) Calculated PEPs and $d$ of a H$_{2}$
penetration into (a) the ideal-PuO$_{2}$(111) surface from the
F1-$Z$ physisorption state; and (b) the reduced-PuO$_{2}$(111) from
T2-$Z$. The total energies of starting geometries with $h_{0}$ =
$4.5$ \r{A} are set to 0 eV and four insets show the atomic
structures of the corresponding reference points. }
\end{figure*}

If H$_{2}$ is barely able to react with PuO$_{2}$(111) surfaces, as
the smallest diatomic molecule, is there any possibility for it to
penetrate vertically into the PuO$_{2}$(111) overlayers? As the
continuations of the vdW-DFT+$U$ PEPs in Fig. 2(a) and 2(b), the
PEPs for H$_{2}$ penetration into ideal- and reduced-PuO$_{2}$(111)
surfaces from the corresponding physisorption states are presented
in\ Fig. 5(a) and 5(b), respectively. We can see from Fig. 5(a) that
the relative energy of the system increases
fleetly during the penetration of H$_{2}$ molecule\ into ideal-PuO$_{2}$%
(111) surface, wherein two insets show atomic structures of two
reference
points (namely, $h=0$ and $-1.0$ \AA ). Since H$_{2}$ touches the surface ($%
h=0$ \AA ), its bond ($d=0.8$ \AA ) is stretched notably, and once
the whole H$_{2}$ enters the surface ($h=-1.0$ \AA ), the H-H bond
is snapped ($d=1.5 $ \AA ) with one H atom binding to the subsurface
Pu-cation and the other segregating out of the surface. Upon that,
the penetration of H$_{2}$ molecule into ideal-PuO$_{2}$(111)
surface is too endothermic to occur under ambient conditions because
the stoichiometric PuO$_{2}$ is so close-grained that the H$_{2}$
needs to overcome a high energy barrier of more than $4$ eV to be
broken before squashing in the PuO$_{2}$ overlayer. For the
penetration of H$_{2}$ molecule into reduced-PuO$_{2}$(111) surface,
we can see from Fig. 5(b) that the PEP curve of T2-$Z$ is stepwise,
namely, the relative energy first increases rapidly during molecular
entry into the surface ($-1.5\leq h\leq 1.3$ \AA ) with the
penetration barrier of $\mathtt{\sim}1.5$ eV, then increases slowly
for the subsurface migration of H$_{2}$ ($-3.0\leq h\leq -1.5$ \AA
), and ultimately increases sharply when H$_{2}$ approaches and
binds to the Pu-cation. Once the H$_{2}$ reacts with the Pu-cation
below, it will dissociate (see the lower inset), and soon the upper
H-atom will be segregated out and stay at the site of surface
O-vacancy binding with three Pu cations nearby (see the upper
inset). Before molecular
dissociation, interestingly, the H-H bond is first stretched slightly while H%
$_{2}$ is hauled step by step into the surface, and then relaxes
back to the equilibrium state after H$_{2}$ entry into the surface,
which differs from the variation of $d$ in Fig. 5(a) and indicates
that just one surface O-vacancy can let the smallest H$_{2}$
molecule enter the PuO$_{2}$ surface layer. But the further
penetration of H$_{2}$ in PuO$_{2}$ overlayer ($h\leq -3.0$ \AA) is
found to be very difficult since H$_{2}$ meets with the
close-grained PuO$_{2}$(111) subsurface layer subsequently.

Thus, according to our current comprehensive study of the
interaction properties between H$_{2}$ molecule and PuO$_{2}$(111)
surfaces, it is found that the close-grained PuO$_{2}$ overlayer can
efficiently prevent the penetration and diffusion of hydrogen (into
the Pu-oxide film), and protect the underlying Pu-metal from
hydrogenation.

\subsection{The interaction between H$_{2}$ and $\protect\alpha $-Pu$_{2}$O$%
_{3}$(111) surface}

In this subsection, we turn to study the interaction mechanisms between H$%
_{2}$ and the stoichiometric $\alpha $-PuO$_{2}$(111) surface. Note
that in addition to the formation of $25$\% oxygen vacancies, there
is an interesting volume expansion ($7.6$\% in theory and $7.4$\% in
experiment) during the PuO$_{2}$¡ú$ŠÁ $-Pu$_{2}$O$_{3}$ isostructure
reduction mainly
due to the strong on-site Coulomb repulsion among the Pu-$5f$ electrons \cite%
{SunPLA2012}. Thus, compared to the close-grained and smooth
PuO$_{2}$(111) surface, the $\alpha $-Pu$_{2}$O$_{3}$(111) seems to
be a porous slab with
 $25$\% native O vacancies located on every O-layer, and its surface is not
smooth any more.

\begin{figure*}[tbp]
\begin{center}
\includegraphics[width=0.8\linewidth]{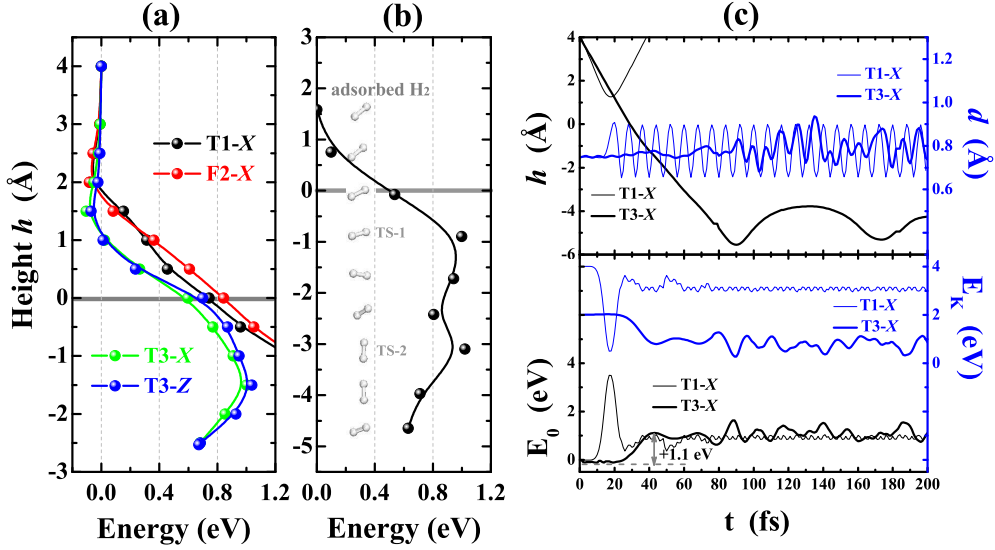}
\end{center}
\caption{(Color online) (a) Calculated PEPs for H$_{2}$ interaction with $\protect\alpha $%
-Pu$_{2}$O$_{3}$(111) surface. (b) Calculated MEP for H$_{2}$
penetration from the physisorption state at T2, with the
orientations of H$_{2}$ during penetration shown by the inset. (c)
AIMD results of H$_{2}$ interaction with $\protect\alpha
$-Pu$_{2}$O$_{3}$(111): $h$ and $d$ of H$_{2}$ (upper panel) and
$E_{\text{0}}$ and $E_{\text{K}}$ of system (lower panel) for
H$_{2}$ dropped from T1-$X$ (with $E_{\text{k}}^{\bot }=4.0$ eV) and
T3-$X$ (with $E_{\text{k}}^{\bot }=2.0$ eV) geometries.}
\end{figure*}

Four representative PEPs of H$_{2}$ molecule approaching $\alpha $-Pu$_{2}$O$%
_{3}$(111) surface are shown in Fig. 6(a). One can see that when H$%
_{2}$ molecule is dropped from T1-$X$ and F2-$X$ geometries, the
corresponding PEPs are similar to those PEPs of H$_{2}$ approaching
ideal-PuO$_{2}$(111)
surface [see Figs. 3(a)-(d)], and thus the T1 and F2 sites on $\alpha $-Pu$%
_{2}$O$_{3}$(111) surface are also very difficult of access. However, when starting from T3-$X$ or T3-$Z$, after the weak physisorption states, H$%
_{2}$ molecule can further be pulled into the $\alpha
$-Pu$_{2}$O$_{3}$(111) overlayer via the surface O-vacancy, which is
similar to the case of H$_{2}$ entry into the reduced-PuO$_{2}$(111)
surface [see Fig. 5(b)]. The energy barrier of H$_{2}$ molecule
penetration into $\alpha $-Pu$_{2}$O$_{3}$ reported by the PEPs is
$\mathtt{\sim}1.1$ eV less than the barrier of $\mathtt{\sim}1.5$ eV
for H$_{2}$ entry the reduced-PuO$_{2}$(111). In order
to accurately obtain the energy barrier and the rate constant $k_{\text{hTST}%
}^{\text{Pen}}$ of H$_{2}$ penetration, we first use the CI-NEB method \cite%
{HENKELMAN2000} with seven images to search out the transition-state
(TS) from the minimum energy pathway (MEP), and then calculate the
vibrational frequencies of H$_{2}$ molecule at the initial
physisorption-state (IS) and the TS. The MEP result and the
orientations of H$_{2}$ molecule during the penetration process are
plotted in Fig. 6(b). One can see that there are two TSs for H$_{2}$
molecule first entry (TS-1) and then diffusion (TS-2) in the $\alpha
$-Pu$_{2}$O$_{3}$ overlayer, respectively. And H$_{2}$ at TS-1
almost horizontally locates $\mathtt{\sim}1.0$ \AA\ below the
surface-layer (with the penetration barrier of $\mathtt{\sim}1.0$ eV) but H$%
_{2}$ at TS-2 tends to be perpendicular to the surface (with the
diffusion barrier of $\mathtt{\sim}0.21$ eV), indicating that
H$_{2}$ seems to be able to rotate freely in the tunnel made of the
vertically distributed O vacancies.

\begin{table}[tbp]
\caption{Calculated normal vibration frequencies $\nu_{\text{n}}$
(in THz, n=$1$-$6$) of H$_{2}$ molecule with the IS, the TS-1 and
the TS-2 structures.}
\label{table1}%
\begin{tabular}{ccccccc}
\hline\hline & \ $\nu_{_{1}}$ \  & \ $\nu_{2}$ \  & \ $\nu_{3}$ \  &
\ $\nu_{4}$ \  & \ $\nu_{5}$ \ & \ $\nu_{6}$ \  \\ \hline
IS & 129.43 & 9.21 & 7.36 & 3.85 & 1.26 & 0.84 \\
TS-1 & 120.64 & 31.34 & 18.14 & 10.85 & 3.82 & 9.14(i) \\
TS-2 & 122.09 & 21.46 & 21.32 & 11.81 & 11.04 & 7.46(i) \\
\hline\hline
\end{tabular}%
\end{table}

Based on the vibrational frequencies of H$_{2}$ listed in Table I,
we can see that H$_{2}$ at the IS vibrates, nearly like a free
molecule, with mainly a symmetric stretching vibration
($\nu_{\text{1}}=129.43$ THz). Whereas, H$_{2}$ at both TSs vibrates
somewhat complicatedly under the influence of $\alpha
$-Pu$_{2}$O$_{3}$ matrix and has one imaginary
frequency. At the room temperature of $T=300$ K, the $k_{\text{hTST}}^{%
\text{Pen}}$ of H$_{2}$ is estimated to be
$\mathtt{\sim}2\times10^{-7}$/s, which is really small mainly due to
the relatively high penetration barrier ($1.0$ eV)
of H$_{2}$ entry the stoichiometric $\alpha $-Pu$_{2}$O$%
_{3}$(111) overlayer. However, in practice the Pu-oxide overlayer on
Pu-metal is really far from being an ideal (defect-free) PuO$_{2}$ or Pu$%
_{2} $O$_{3}$, and other defects such as microcracks or some forms
of impurity inclusions in the overlayer can prominently promote the
probability of H$_{2} $ penetration. Therefore, it is certainly one
challenging task to predict and control the length of induction
period in Pu-hydrogenation. Here, at the atomic and molecular level,
we just focus on understanding the different roles of ideal
PuO$_{2}$ and $\alpha $-Pu$_{2}$O$_{3}$ overlayers in
Pu-hydrogenation, and at this point our theoretical results have
well
revealed the major difference between PuO$_{2}$ and $\alpha $-Pu$_{2}$O$_{3}$%
, namely, PuO$_{2}$ is airtight but $\alpha $-Pu$_{2}$O$_{3}$ is
porous for the smallest H$_{2}$ molecule.

\begin{figure*}[tbp]
\begin{center}
\includegraphics[width=0.85\linewidth]{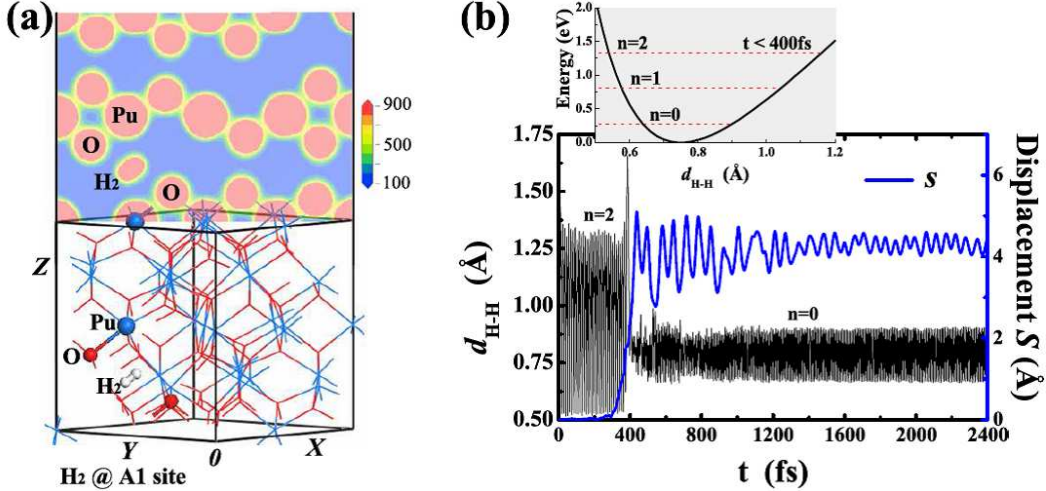}
\end{center}
\caption{ (Color online) (a) Section plane of total electron-density
(top) and atomic structure (bottom) of
$\protect\alpha$-Pu$_{2}$O$_{3}$ matrix with a H$_{2}$ molecule at
A$_{1}$ site. (b) NVE-AIMD
evolutions of $d$ and displacement $s$ of a H$_{2}$ molecule with internal vibrational excitation ($n=2$ as shown by the inset) in $\protect%
\alpha$-Pu$_{2}$O$_{3}$.}
\end{figure*}

Furthermore, the AIMD simulations are also performed to depict the
interaction dynamics for H$_{2}$ impingement against the $\alpha $-PuO$_{2}$%
(111) surface. Figure 6(c) shows the representative NVE-AIMD
trajectories started from the T1-$X$ and T3-$X$ with
$h_{\text{0}}=4$ \AA , where the evolutions of height $h$ and
bond-length $d$ of H$_{2}$ (upper panel),
the relative electronic free-energy $E_{\text{0}}$, and the kinetic energy $E_{\text{k%
}}$ of the system (lower panel) are shown. We can see that even when
the initial $E_{\text{k}}^{\bot }$ reaches $4.0$ eV, H$_{2}$ seems
to be unable to
touch the T1 site on $\alpha $-Pu$_{2}$O$_{3}$(111) surface ($h\geq 1.2$ \AA %
) and soon bounces back with the zero-point vibration, which is very
unlike the dissociation of H$_{2}$ (with $E_{\text{k}}^{\bot }$ of
$4.0$ eV) on reduced-PuO$_{2}$(111) surface [see Fig. 4(f)] but is
like the diffraction behavior on ideal-PuO$_{2}$(111) surface [see
Fig. 3(f)]. The reasons are that (i) the $\alpha
$-Pu$_{2}$O$_{3}$(111) surface is not smooth with many protrusions
such as the T1 site, and actually H$_{2}$ can contact the O-anion at
T1 site; and (ii) on the other hand it will take much more energy
to dig an O-anion out of the surface since further reduction of $\alpha $-Pu$%
_{2}$O$_{3}$ is very difficult \cite{SunPLA2012}. However, the
H$_{2}$ molecule with an initial $E_{\text{k}}^{\bot }$ of $2.0$ eV
can stride over an energy barrier
of $\mathtt{\sim}1.1$ eV to penetrate via T3 site into the $\alpha $-Pu$_{2}$O%
$_{3}$(111) overlayer, and then diffuses easily along the devious
alleyway made of the continuous distributed O vacancies. During the
H$_{2}$ penetration and diffusion, the internal vibration of H$_{2}$
molecule is fitful and is especially so after the first collision
with Pu-cation ($t=90$
fs). Although the inhalation of a H$_{2}$ molecule in the $\alpha $-Pu$%
_{2}$O$_{3}$(111) slab is endothermic (see $E_{\text{0}}$ of the
system), the hot H$_{2}$ molecule can release the redundant kinetic
energy and change the direction of movement via successive
collisions with $\alpha $-Pu$_{2}$O$_{3}$ matrix,
and as a result H$_{2}$ seems to be neither escape out of nor dissociate in $\alpha $%
-Pu$_{2}$O$_{3}$(111) during the simulation process of $3$ ps.

\begin{table*}[tbp]
\caption{Calculated bond-length $d$ (in \AA ), adsorption energy $E_{%
\text{ad}}$ (in eV), and stretching vibration frequencies
$\nu_{\text{1}}$ (in THz) of H$_{2}$ molecule at four adsorption
states A$_{\text{n}}$ and four
transition states T$_{\text{n}}$ (n=1-4). Calculated room-temperature (300 K) rate constant $k_{\text{%
hTST}}^{\text{Diff}}$ (in s$^{-1}$) of the bilateral diffusion of
H$_{2}$ in
$\protect\alpha $-Pu$_{2}$O$_{3}$ is also listed.}%
\begin{tabular} {ccccccccc}
\hline\hline & \ A$_{_{1}}$ \  & \ T$_{_{1}}$ \  & \ A$_{2}$ \  & \
T$_{2}$ \  & A$_{3}$ & T$_{3}$ & A$_{4}$ & T$_{4}$ \\ \hline
$d$ & 0.760 & 0.751 & 0.760 & 0.764 & 0.765 & 0.758 & 0.767 & 0.765 \\
$E_{\text{ad}}$ & 0.665 & 0.925 & 0.669 & 1.125 & 0.606 & 0.845 &
0.601 &
1.134 \\
$\nu_{\text{1}}$ & 120.84 & 124.27 & 120.70 & 118.38 & 118.88 &
121.30 & 118.21
& 116.84 \\
$k_{\text{hTST}}^{\text{Diff}}$ & $\rightarrow $ & 1.75$\times $10$^{9}$ & $%
\rightarrow $ & 2.81$\times $10$^{4}$ & $\rightarrow $ & 5.16$\times $10$%
^{9} $ & $\rightarrow $ & 7.22$\times $10$^{3}$ \\
$k_{\text{hTST}}^{\text{Diff}}$ & $\leftarrow $ & 2.18$\times $10$^{9}$ & $%
\leftarrow $ & 4.94$\times $10$^{3}$ & $\leftarrow $ & 3.21$\times
$10$^{9}$ & $\leftarrow $ & 1.90$\times $10$^{4}$ \\ \hline\hline
\end{tabular}%
\end{table*}

In order to gain detailed insights into the adsorption states and
the
diffusion behaviors of H$_{2}$ molecule in the $\alpha $-Pu$_{2}$O$_{3}$%
(111) matrix, we first carry out systematic total-energy
calculations to search for favorable sites for H$_{2}$ to stay at.
Then, based on the AIMD simulations within canonical ensemble
(NVT-AIMD), we discuss the
temperature effect on the state or behavior of H$_{2}$ in $\alpha $-Pu$_{2}$O%
$_{3}$. Figure 8(a) shows four relatively favorable molecular
mass-center sites A$_{n}$ ($n$=$1$-$4$) for H$_{2}$ adsorption, and
the adsorption energies of H$_{2}$ at these four sites are listed in
Table II, where H$_{2}$ molecule can rotates slightly around the
four sites and the A$_{4}$ site seems to be a little more favorable
with a positive (endothermic) adsorption energy $E_{\text{ad}}$ of
$+0.601$ eV. Subsequent NVT-AIMD simulations at three temperatures
($T=300$, $600$, and $900$ K) show that the molecules just moves
around their initial adsorption sites with the zero-point vibration
and do not diffuse or dissociate within the isothermal period of $5$
ps. The atomic adsorption energy is calculated to be endothermic
($+0.69$ eV/atom), which indicates that the dissociative adsorption
of H$_{2}$ in $\alpha $-Pu$_{2}$O$_{3}$ should be more endothermic
than the molecular adsorption. Thus, in contrary to the slightly
exothermic adsorption of H atom to form OH group in bulk PuO$_{2}$,
H$_{2}$ molecule and even H atom can not easily react with the
$\alpha $-Pu$_{2}$O$_{3}$ matrix since the reduction of $\alpha
$-Pu$_{2}$O$_{3}$ is much more difficult than that of PuO$_{2}$
\cite{SunPLA2012}. The exothermic adsorption of H atom in bulk
PuO$_{2}$ is also found to be similar to that of H atom in CeO$_{2}$
\cite{SOHLBERG2001}, which can reduce the $+4$ cation in both
fluorite-structured oxides.

\begin{figure*}[tbp]
\begin{center}
\includegraphics[width=0.75\linewidth]{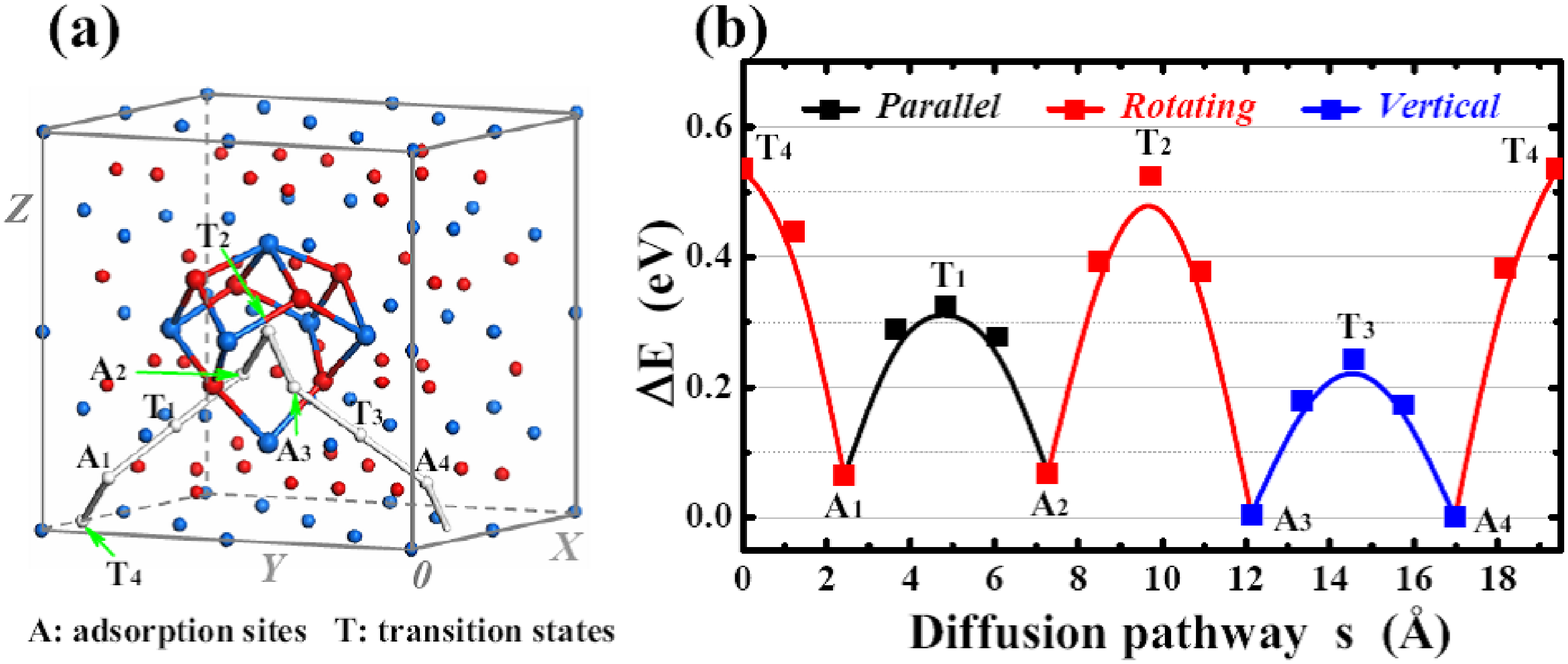}
\end{center}
\caption{ (Color online) (a) Four adsorption sites A$_{1}$-A$%
_{4}$ and four transition-state sites T$_{1}$-T$_{4}$ of H$_{2}$ in $\protect%
\alpha $-Pu$_{2}$O$_{3}$, which are linked with thick white lines to
indicate the whole diffusion path of H$_{2}$. (b) MEPs between
adjacent adsorption sites, with the A$_{4}$-site adsorption state as
the reference point ($E_{0}=0$ eV). Lines of different colors are
used to guide the eyes.}
\end{figure*}

According to Table II, the adsorption properties of H$_{2}$
molecules at four sites seem to be close to each other, and their
temperature-dependent behaviors in $\alpha $-Pu$_{2}$O$_{3}$ are
also very similar. For the convenience of depiction, we just plot in
Fig. 7(a) the atomic- and electronic-structure results of the
A$_{1}$-site adsorption system. We can see that the H$_{2}$ molecule
actually stays in suspension at A$_{1}$ site, which is near the
location of an O-anion in the
isostructure PuO$_{2}$ supercell, i.e., the native O-vacancy site in $%
\alpha $-Pu$_{2}$O$_{3}$. The charge density distribution in Fig.
7(a) indicates that there is no evidence of chemical binding of
H$_{2}$ molecule to the nearby Pu-cation or O-anion, and there seems
to be enough space for the movement or diffusion of H$_{2}$ molecule
in the $\alpha $-Pu$_{2}$O$_{3}$ matrix. Furthermore, in the
NVE-AIMD, we first consider the internal vibrational excitation of
H$_{2}$ molecule at A$_{1}$ site [with $n=2$ in Eq. (3)] and then
let the system relax freely, as is shown in Fig. 7(b). We find that
during the initial $400$ fs, the H$_{2}$ vibrates drastically at
A$_{1}$ site. Then, this hot H$_{2}$ molecule touches the nearby
Pu-cation at $t=400$ fs, and meanwhile transforms most of the
vibration energy into the translation and
rotation energy. Soon after that, H$_{2}$ migrates quickly to the A$%
_{2}$ site and transfers its kinetic energy towards the vibration of the $%
\alpha $-Pu$_{2}$O$_{3}$ matrix via the successive collisions with
the matrix within $\mathtt{\sim}400$ fs. Eventually ($t\geq 800$
fs), the H$_{2}$ molecule at A$_{2}$ site comes back to the
zero-point vibration
state after releasing most of its kinetic energy. The displacement $S$ of H$%
_{2}$ as a function time ($t\leq 2.4$ ps) plotted in Fig. 7(b)
indicates that the local movement (A$_{1}\rightarrow $A$_{2}$) of
H$_{2}$ is easy but the long-distance diffusion through the four
sites seems to be very time-consuming, which can not be directly
studied by the AIMD simulations. Overall, this interesting example
reveals that the H$_{2}$ molecule tends to keep its ground state
with the zero-point vibration.

Figure 8(a) shows that the four sites in a crystal cell of $\alpha $-Pu$_{2}$O$%
_{3}$ are actually connectible by the O-vacancy distributed continuously in the $%
\alpha $-Pu$_{2}$O$_{3}$ matrix, so that if given enough time,
H$_{2}$ can penetrate through the $\alpha $-Pu$_{2}$O$_{3}$
overlayer and the diffusion of H$_{2}$ in $\alpha $-Pu$_{2}$O$_{3}$
seems to be continuous in a stepwise way. By using the CI-NEB
method, we search out four TSs [with their sites labeled by T$_{n}$
($n$=$1$-$4$) in Fig. 8(a)] from the calculated minimum energy
pathways (MEPs) between adjacent sites, which are plotted in Fig.
8(b). We can see from Fig. 8 that (i) the whole diffusion pathway is
a close
loop and consists of four short routes in the periodical $\alpha $-Pu$_{2}$O$%
_{3}$ matrix; and (ii) when diffusing along the linear routes (from
A$_{1}$ to A$_{2}$ and from A$_{3}$ to A$_{4}$, and vice versa), the
H$_{2}$ can keep parallel and vertical to the routes and the
corresponding lowest diffusion barriers are $0.26$ eV
(A$_{1}\rightarrow $A$_{2}$) and $0.24$ eV (A$_{3}\rightarrow
$A$_{4}$); whereas (iii) when migrating along the
meandering routes (A$_{4}$ $\rightleftarrows $ A$_{1}$ and A$_{2}$ $%
\rightleftarrows $ A$_{3}$). The corresponding rightward diffusion
barriers are doubled to be $0.53$ and $0.46$ eV because H$_{2}$
needs to rotate during the migration courses and swerve at the
transition states (T$_{2}$
and T$_{4}$). Furthermore, we calculate the vibrational frequencies of H$%
_{2} $ at four adsorption sites and four transition states, with the
major stretching vibration frequencies listed in Table II. Then,
following Eq.
(4), we figure out the rate constant ($k_{\text{hTST}}^{\text{Diff}}$) of H$%
_{2}$ diffusion in bulk $\alpha $-Pu$_{2}$O$_{3}$ and list the
calculated results of $k_{\text{hTST}}^{\text{Diff}}$ at room
temperature ($T=300$ K) in Table II. We can see that the
$k_{\text{hTST}}^{\text{Diff}}$ of H$_{2}$ diffusion from a
metastable state to a relative stable state is always larger than
the $k_{\text{hTST}}^{\text{Diff}}$ of the reverse diffusion,
and the probabilities for H$_{2}$ diffusion along linear routes (A$%
_{1}$ $\rightleftarrows $ A$_{2}$ and A$_{3}$ $\rightleftarrows $
A$_{4}$) are about five or six orders of magnitude larger than the
case of H$_{2}$ migration along meandering routes, indicating that
the diffusion kinetics of H$_{2}$ in $\alpha $-Pu$_{2}$O$_{3}$ is
mainly determined by the thermodynamical parameter diffusion
barrier.

Assuming sequential first-order kinetics, the overall diffusion rate
for H$_{2}$ in the $\alpha $-Pu$_{2}$O$_{3}$
crystal cell can be simply estimated as $\frac{1}{k_{\text{overall}}^{\text{%
Diff}}}=\sum_{i=1}^{4}\frac{1}{k_{i}^{\text{Diff}}}$, where $k_{i}^{\text{Diff%
}}$ is the rate constant of the $i$th step diffusion. At $300$ K,
the rightward (A$_{1}\mathtt{\rightarrow}$A$_{4}$)
$k_{\text{overall}}^{\text{Diff}}$ of H$_{2}$ in $\alpha
$-Pu$_{2}$O$_{3}$ is $\mathtt{\sim}5.7\times 10^{3}$/s, and the
leftward one is $\mathtt{\sim}3.9\times 10^{3}$/s, indicating that
the H$_{2}$ migration along meandering routes is the
rate-determining step for the diffusion kinetics of H$_{2}$ in
$\alpha $-Pu$_{2}$O$_{3}$. Considering the small
$k_{\text{hTST}}^{\text{Pen}}$ of $\mathtt{\sim}2\times 10^{-7}$/s
for H$_{2}$ penetration into $\alpha $-Pu$_{2}$O$_{3}$(111)
overlayer, the diffusion of H$_{2}$ in bulk $\alpha
$-Pu$_{2}$O$_{3}$ seems to be much easier\ since the
$k_{\text{overall}}^{\text{Diff}}$ is at least ten orders of
magnitude larger than the $k_{\text{hTST}}^{\text{Pen}}$.
Thus, under oxygen-poor conditions, the course of H$_{2}$ entry the $\alpha $%
-Pu$_{2}$O$_{3}$ overlayer is the final rate-determining step for
H$_{2}$ penetration through the $\alpha $-Pu$_{2}$O$_{3}$ to reach
the Pu-metal. In order to quantificationally determine the rate
constants of H$_{2}$ penetration, the practical surface
configuration of $\alpha $-Pu$_{2}$O$_{3}$ overlayer is a critical
factor. But the complex surface conditions of oxide-coated Pu are
undoubtedly not as simple as the stoichiometric $\alpha
$-Pu$_{2}$O$_{3}$(111) surface considered in this work, so that we
just take the first step towards understanding the hydrogenation of
Pu-oxide coated Pu metal.

\section{CONCLUSIONS}

In summary, based on the vdW-DFT+$U$ scheme, we have explored the
different roles of the Pu-oxide overlayers in the hydrogenation of
Pu underlayer. Three model surfaces of Pu-oxides are considered by
varying oxygen conditions. We have found that the physisorption
state of H$_{2}$ on PuO$_{2}$(111) surfaces can not be captured by
pure DFT+$U$ calculations until the van der Waals correction is
taken into account. We have shown that the surface O-vacancy induces
a local polarization and produce a stronger van der Waals attraction
for H$_{2}$, thus the physisorption energy of H$_{2}$ near the
O-vacancy on the reduced PuO$_{2}$(111) surface becomes higher by
$\mathtt{\sim}70$ meV than that on the ideal PuO$_{2}$(111) surface.
However, the physisorbed H$_{2}$ molecules can not further get close
to PuO$_{2}$(111) surfaces. In agreement with the static PEP
results, our AIMD simulations have shown that the H$_{2}$ molecule
will bounce back to the vacuum when its initial kinetic energy is
not high enough. Until the initial $E_{\text{k}}^{\bot }$ reaches
$\mathtt{\sim}5.0$ eV ($\mathtt{\sim}4.0$ eV), the H$_{2}$ molecule
parallel to stoichiometric (reduced) PuO$_{2}$(111) surface can be
broken. Interestingly, these collision-induced dissociation
behaviors of H$_{2}$ have been found to be very site-specific. For
instance, two H atoms can chemically bind with two O-anions to form
hydroxyl groups and can also react with one O-anion to produce a
H$_{2}$O molecule. Although the dissociative adsorption of H$_{2}$
is very exothermic and can reduce PuO$_{2}$(111) surfaces,
it should be an extremely rare event due to the high energy barriers ($3.2$ and $%
2.0$ eV for H$_{2}$ on the ideal- and reduced-PuO$_{2}$(111)
surfaces, respectively). Therefore, the close-grained PuO$_{2}$
overlayer acts as a diffusion barrier to control the permeation of
hydrogen and lengthen the initiation time of Pu-hydrogenation.

As a product of isostructure reduction of PuO$_{2}$, $\alpha $-Pu$_{2}$O$%
_{3}$ has $25$\% oxygen vacancies located along $\langle
$111$\rangle $ diagonals and thus seems to be a porous slab. Our
AIMD study have directly revealed that the H$_{2}$ molecule can
overcome an barrier of $\mathtt{\sim}1.0$ eV
and directly penetrate into the $\alpha $-Pu$%
_{2}$O$_{3}$(111) film via the O vacancies. By examining the
temperature effect and the internal vibrational excitations of
H$_{2}$, we have also provided a detailed insight into the
interaction dynamics between H$_{2}$ and $\alpha $-Pu$_{2}$O$_{3}$.
In $\alpha $-Pu$_{2}$O$_{3}$, the hot H$_{2}$ molecule prefers to
release its energy via successive collisions and come back to its
ground state with zero-point vibration. The diffusion behavior of
H$_{2}$ in $\alpha$-Pu$_{2}$O$_{3}$ is also systematically
investigated by searching out the minimum diffusion paths and
calculating the diffusion rate constants of H$_{2}$. Our results are
consistent with the general experimental observations, and come to
the conclusion that the PuO$_{2}$ overlayer can hinder the hydrogen
penetration, provided the isostructure reduction of PuO$_{2}$ to
$\alpha $-Pu$_{2}$O$_{3}$ is efficiently suppressed.

\begin{acknowledgements}

This work was supported by the Foundations for Development of
Science and Technology of China Academy of Engineering Physics under
Grants Nos. 2010B0301048, 2011A0301016, and 909-07, and partially
supported by the NSFC under Grants No. 11004012, No. 51071032, No.
11205019. and No. 11275032. \

\end{acknowledgements}

\end{document}